\documentclass[aps,pra,twocolumn]{revtex4}
\usepackage{amsfonts}
\usepackage{amsmath}
\usepackage{amssymb}
\usepackage{epsfig}
\usepackage{bm}

\def\be{\begin{equation}}
\def\ee{\end{equation}}
\def\bea{\begin{eqnarray}}
\def\eea{\end{eqnarray}}

\begin{document}

\title{Superradiant control of $\gamma $-ray propagation by vibrating
nuclear arrays}
\author{Xiwen Zhang\footnote{%
E-mail: xiwen@physics.tamu.edu} and Anatoly A. Svidzinsky}
\affiliation{Department of Physics and Astronomy, Texas A\&M University, College Station,
Texas 77843, USA}
\date{\today }

\begin{abstract}
The collective nature of light interactions with atomic and nuclear ensembles
yields the fascinating phenomena of superradiance and radiation trapping. We
study the interaction of $\gamma $ rays with a coherently vibrating periodic array
of two-level nuclei. Such nuclear motion can be generated, e.g., in ionic
crystals illuminated by a strong driving optical laser field. We find that
deflection of the incident $\gamma $ beam into the Bragg angle can be switched
on and off by nuclear vibrations on a superradiant time scale determined by
the collective nuclear frequency $\Omega _{a}$, which is of the order of
terahertz. Namely, if the incident $\gamma $ wave is detuned from the
nuclear transition by frequency $\Delta \gg \Omega _{a}$ it passes through
the static nuclear array. However, if the nuclei vibrate with frequency $%
\Delta $ then parametric resonance can yield energy transfer into the Bragg
deflected beam on the superradiant time scale, which can be used for fast
control of $\gamma$ rays.
\end{abstract}

\maketitle

\section{Introduction}

Gamma rays are widely used in contemporary technologies for material
modification, food sterilization and testing for weak points in welded
structures. Medical applications of $\gamma$ rays include the imaging technique
of positron emission tomography and radiation therapies to treat cancerous
tumors as well as detecting brain and cardiovascular abnormalities.

Since the discovery of recoilless nuclear resonance by M\"{o}ssbauer \cite%
{Mossbauer58, Craig59}, studies of the interaction between $\gamma $ rays
and M\"{o}ssbauer nuclear ensembles have undergone rapid development and have
yielded many real and potential applications in, e.g., M\"{o}ssbauer
spectroscopy \cite{Dyar06} and quantum information~\cite{Shvydko96,Palffy09}%
. Due to their small wavelength, $\gamma $ rays are naturally suitable for
achieving high spatial resolution and for making small quantum photonic
circuits \cite{Liao12}.

However, control of $\gamma $ rays still remains a challenging problem.
Coherent effects, such as level mixing induced transparency \cite%
{Coussement02}, electromagnetically induced transparency~\cite{Rohlsberger12}%
, $\gamma $ echo \cite{Helisto91,Smirnov96}, phase modulation \cite{Shakhmuratov11,Shakhmuratov13}, and the
nuclear lighthouse effect \cite{Rohlsberger00}, can be adopted to
manipulate $\gamma $ radiation. Modulation of M\"{o}ssbauer radiation by
pulsed laser excitation was demonstrated in~\cite{Vagizov12}. The total
reflection of the grazing incidence was used to reflect $\gamma $ rays, but
application of this technique is limited due to small grazing angle.
Development of the $\gamma $-ray optics led to the design of the Laue lens \cite%
{Barriere09} via nuclear Bragg diffraction~\cite{Gerdau85}. It has also been
suggested that $\gamma $ rays can be manipulated using Delbr\"{u}ck
scattering \cite{Habs12}.

Effective control of $\gamma $ rays requires further advancements and
innovations. Development of a fast switch of $\gamma $ rays is important for
extending the time resolution of $\gamma $-ray sources and for increasing
the operating speed of $\gamma $-ray quantum information processing.
Nanosecond $\gamma $-ray switching has been realized by magnetically
manipulating nuclear excitation based on the quantum beat in nuclear Bragg
scattering \cite{Shvydko94}. Picosecond x-ray Bragg switch utilizing laser-generated phonons was proposed \cite{Bucksbaum99} and later demonstrated
experimentally \cite{DeCamp01,Herzog10}.

In this paper we investigate a way to control propagation of a $\gamma $-ray
beam through a crystal by controlling its collective absorption and
reemission by many nuclei. Collective spontaneous emission from atomic
ensembles has been a subject of long-standing interest since the pioneering
work of Dicke \cite{Dicke54}. The collective nature of light interaction yields
fascinating effects such as superradiance and radiation trapping even at the
single-photon level. Recent studies focus on collective, virtual and
nonlocal effects in such systems \cite%
{Scully06,Eberly06,Mazets07,Svid08,Svid08a,Frie08a,Frie08,Porras08,Scully09Science,Pedersen09,Svid09,Svid10a,Friedberg10,Svid10,Berman11,Svid12,Ji07,ZhangXF13}%
. The Josephson effect for photons in two weakly linked microcavities is an
example of the collective physics in coupled atom-cavity systems~\cite{Ji09}.

The interaction of light with ordered arrays of nuclei in crystals offers new
perspectives. For example, a photon collectively absorbed by a random medium
(e.g., gas) will be reemitted in the same direction as the incident photon
\cite{Scully06}. However, in the case of a crystal lattice, collective
reemission can occur in several directions (Bragg angles). The interaction
strength between the $\gamma $-ray beam and the crystal depends on the
detuning $\Delta $ of the photon frequency from the nuclear transition. Here
we show that one can redirect a $\gamma $-ray beam into a desirable Bragg
angle by making the crystal lattice coherently vibrate with frequency $%
\Delta $ which lies, e.g., in the infrared region. Such lattice vibrations
are in the combination parametric resonance with the frequency difference
between two eigenmodes of the coupled light-nuclear system which results in
resonant energy transfer from the incident $\gamma $-ray beam to the wave
propagating at the Bragg angle. This process is analogous to the parametric
frequency mixing in propagating circuits~\cite{Tien58}.

Nuclear vibrations can be generated by a driving laser pulse and can be
turned on and off on a short time scale. $\gamma $-ray redirection, produced
by parametric resonance, occurs on a time scale determined by the collective
nuclear frequency $\Omega _{a}$ which typically lies in the terahertz region. This
mechanism allows us to control propagation of high frequency $\gamma $
photons by driving the system, e.g., with an infrared laser.

\section{The model and derivation of basic equations}

\label{sec:formulism}

\begin{figure}[h]
\begin{center}
\epsfig{figure=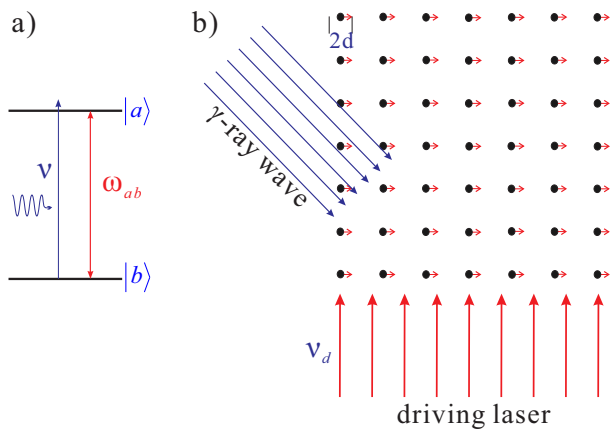, width=8cm}
\end{center}
\caption{(Color online) Illustration of the model. (a) Energy diagram of the
two-level nuclear system. (b) Present model: an incident $\protect\gamma $%
-ray plane wave interacts collectively with a recoilless nuclear array,
while the strong optical laser field produces coherent oscillations of the
nuclei with amplitude $d$ and frequency $\protect\nu _{d}$.}
\label{Setup}
\end{figure}

We consider a perfect crystal composed of two-level ($a$ and $b$)\ nuclei
with transition frequency $\omega _{ab}$ as shown in Fig. \ref{Setup}(a).
The nuclear transition frequency $\omega _{ab}$ typically lies in the hard x-ray
or $\gamma $-ray region. Nuclei are located at positions $\mathbf{r}_{j}$
and form a periodic lattice, where the index $j$ labels different nuclei.
Typically, the inter nuclei spacing is much larger than the nuclear
radiation wave length $\lambda _{ab}=2\pi c/\omega _{ab}$, where $c$ is the
speed of light.

We assume that the lattice is coherently excited so that nuclei oscillate along
the direction given by a unit vector $\hat{n}$ around their equilibrium
positions $\mathbf{r}_{j}^{0}$. The oscillation frequency $\nu _{d}$ lies in the
infrared or visible region. In ionic crystals such oscillations can be produced, e.g., by a strong linearly polarized driving laser pulse with
frequency $\nu _{d}$. A typical example is potassium iodide crystal, which
has a face-centered-cubic unit cell of iodide ions with potassium ions in
octahedral holes. By applying an external driving field one can make ions K$%
^{+}$ and I$^{-}$ move in opposite directions such that nuclei of the
same species will oscillate in unison. Both K and I have M\"{o}ssbauer
isotopes. Namely, $^{40}$K has a M\"{o}ssbauer transition with energy $29.8$
keV and spontaneous decay rate $\Gamma =2.4\times 10^{8}$ s$^{-1}$, while $%
^{127}$I has a transition with energy $58.6$ keV and $\Gamma =5.1\times 10^{8}$
s$^{-1}$.

We consider an interaction of high-frequency (x- or $\gamma $-ray) photons with
a coherently vibrating nuclear lattice of a particular M\"{o}ssbauer isotope.
The presence of nuclei of another species in the crystal is irrelevant since
they have a very different transition frequency. We assume that the motion of each
nuclei $j$ involved in the interaction is given by
\begin{equation}
\mathbf{r}_{j}(t)=\mathbf{r}_{j}^{0}+\hat{n}f(t),
\end{equation}%
where%
\begin{equation}
f(t)=d\sin (\nu _{d}t).
\end{equation}%
Here $\nu _{d}\ll \omega _{ab}$ and $d\lesssim \lambda _{ab}$ is the
amplitude of the laser induced nuclei oscillations.

In our model a weak, plane, linearly polarized $\gamma $-ray wave with the
wave vector $\mathbf{k}_{1}$ and frequency $\nu _{1}=ck_{1}$ detuned from
the nuclear transition frequency $\omega _{ab}$ by an amount $\Delta _{1}\ll
\omega _{ab}$ enters the crystal and collectively interacts with the
oscillating recoilless nuclei [see Fig. \ref{Setup}(b)]. For the sake of
simplicity, we consider only the interaction of the wave with the nuclei and
disregard interaction with electrons. Processes such as internal conversion,
the photoelectric effect \cite{Borobchenko69}, and electron Rayleigh scattering \cite%
{Black64,Stepanov74} are neglected.

We treat the problem in a semiclassical formalism. Namely, the electromagnetic
field $E(t,\mathbf{r})$ of the $\gamma $ ray is described by the classical
Maxwell equation%
\begin{equation}
\left( \nabla ^{2}-\frac{1}{c^{2}}\frac{\partial ^{2}}{\partial t^{2}}%
\right) E=\mu _{0}\frac{\partial ^{2}P}{\partial t^{2}}
\end{equation}%
in which $\mu _{0}$ is the permeability of free space and the polarization of
the medium
\begin{equation}
P=\sum_{j}\left( d_{ba}\rho _{ab}^{\text{j}}+\text{c.c.}\right) \delta \left(
\mathbf{r}-\mathbf{r}_{j}(t)\right)  \label{p1}
\end{equation}%
is determined by the off-diagonal elements of the nuclear density matrix $%
\rho _{ab}^{\text{j}}$. In Eq. (\ref{p1}) the summation is taken over nuclei
that are treated as point particles located at positions $\mathbf{r}_{j}(t)$%
. Assuming that the nuclear transition matrix element $d_{ab}$ is real and
introducing the Rabi frequency of the $\gamma $-ray field $\Omega _{\gamma }(t,%
\mathbf{r})=d_{ab}E(t,\mathbf{r})/\hbar $, we obtain%
\begin{align}
& \left( c^{2}\nabla ^{2}-\frac{\partial ^{2}}{\partial t^{2}}\right) \Omega
_{\gamma }(t,\mathbf{r})  \notag \\
& =\frac{c^{2}\mu _{0}|d_{ab}|^{2}}{\hbar }\frac{\partial ^{2}}{\partial
t^{2}}\sum_{j}\left( \rho _{ab}^{j}+c.c.\right) \delta (\mathbf{r}-\mathbf{r}%
_{j}(t))\,.  \label{m1}
\end{align}%
Equation (\ref{m1}) must be supplemented by the evolution equation for the
nuclear density matrix%
\begin{equation}
\frac{\partial \rho _{ab}^{j}(t)}{\partial t}=-i\omega _{ab}\rho
_{ab}^{j}(t)+i\Omega _{\gamma }(t,\mathbf{r}_{j}(t))(1-2\rho _{aa}^{j})\,.
\label{c1}
\end{equation}%
We assume that nuclear excitation remains weak, so the population of the
excited state $\rho _{aa}^{j}$ can be disregarded. We look for a solution in
the form%
\begin{equation}
\Omega _{\gamma }(t,\mathbf{r})=\Omega (t,\mathbf{r})e^{-i\omega
_{ab}t}+c.c.,
\end{equation}%
\begin{equation}
\rho _{ab}^{j}(t)=\rho ^{j}(t)e^{-i\omega _{ab}t},
\end{equation}%
where $\Omega (t,\mathbf{r})$ and $\rho ^{j}(t)$ are slowly varying
functions of $t$ as compared to the fast oscillating exponentials. In the
slowly varying amplitude approximation, Eqs. (\ref{m1}) and (\ref{c1})
reduce to

\begin{align}
\bigg\{ &\frac{\partial }{\partial t}+\frac{c^{2}}{2i\omega _{ab}}\left[
\left( \frac{\omega _{ab}}{c}\right) ^{2}+\nabla ^{2}\right] \bigg\} \Omega
(t,\mathbf{r}) \notag \\
&=i\frac{\Omega _{a}^{2}}{N}\sum_{j}\rho ^{j}(t)\delta (\mathbf{r}-\mathbf{r}%
_{j}(t))\,,  \label{s1}
\end{align}
\begin{equation}
\frac{\partial \rho ^{j}(t)}{\partial t}=i\Omega (t,\mathbf{r}_{j}(t))\,,
\label{x0}
\end{equation}%
where
\begin{equation}
\Omega _{a}=\sqrt{\frac{c^{2}\mu _{0}|d_{ab}|^{2}\omega _{ab}N}{2\hbar }}=%
\sqrt{\frac{3cN\lambda _{ab}^{2}\Gamma }{8\pi }}  \label{y0}
\end{equation}%
is the collective nuclei frequency proportional to the square root of the
average nuclei density $N$ and $\Gamma $ is the spontaneous decay rate of
the nuclear transition. Physically, $\Omega _{a}$ determines the time scale
of the collective resonant absorption of the incident photon by the medium
\cite{Burnham69,Svid08a,Svid12} and typically is of the order of terahertz.
For example, for a $29.8$-keV transition of the $^{40}$K M\"{o}ssbauer isotope
that spontaneously decays at the rate $\Gamma =2.4\times 10^{8}$ s$^{-1}$
if we take the nuclei density to be $N=8\times 10^{21}$~cm$^{-3}$ we obtain $%
\Omega _{a}\sim 3\times 10^{11}$~s$^{-1}$.

A crystal is made up of a periodic arrangement of atoms (Bravais lattice)
that form an infinite array of discrete points given by $\mathbf{r}%
_{j}^{0}=m_{1}\mathbf{a}_{1}+m_{2}\mathbf{a}_{2}+m_{3}\mathbf{a}_{3}$, where
$m_{i}$ ($i=1,2,3$) are any integers and $\mathbf{a}_{i}$ are the primitive
lattice vectors. As a consequence, $\sum_{j}\delta \left( \mathbf{r}-\mathbf{%
r}_{j}(t)\right) $ is a periodic function of $\mathbf{r}$ with periods $%
\mathbf{a}_{i}$ and, thus, it can be expanded in the Fourier series as%
\begin{align}
\sum_{j}\delta \left( \mathbf{r}-\mathbf{r}_{j}(t)\right) &=N\sum_{m}e^{i
\mathbf{K}_{m}\cdot \left[ \mathbf{r-r}_{j}(t)\right] } \notag \\
&=N\sum_{m}e^{i\mathbf{K}_{m}\cdot \left[ \mathbf{r-}\hat{n}f(t)\right] }\,,  \label{s}
\end{align}%
where $\mathbf{K}_{m}=m_{1}\mathbf{b}_{1}+m_{2}\mathbf{b}_{2}+m_{3}\mathbf{b}%
_{3}$, $\mathbf{b}_{1,2,3}$ are the primitive vectors of the reciprocal
lattice and $N$ is the average nuclear density.

\begin{figure}[h]
\begin{center}
\epsfig{figure=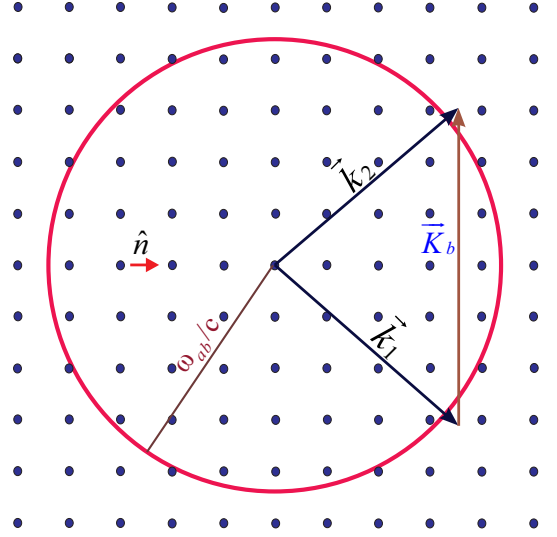, width=7cm}
\end{center}
\caption{(Color online) Two dimensional reciprocal lattice of the crystal is
shown by dots. The incident $\protect\gamma $-ray beam with the wave vector $%
\mathbf{k}_{1}$ is detuned from the nuclear transition frequency $\protect%
\omega _{ab}$. The incident wave is coupled with the Bragg wave that has
wave vector $\mathbf{k}_{2}=\mathbf{k}_{1}+\mathbf{K}_{b}$, where $\mathbf{K}%
_{b}$ is a reciprocal lattice vector.}
\label{Reciprocal lattice}
\end{figure}

We look for $\rho ^{\text{j}}(t)$ in the form
\begin{equation}
\rho ^{\text{j}}(t)=\rho (t)e^{i\mathbf{k}_{1}\mathbf{\cdot r}_{j}^{0}}.
\end{equation}%
Multiplying both sides of Eq. (\ref{s}) by $e^{i\mathbf{k}_{1}\cdot \mathbf{r%
}}$ we obtain%
\begin{equation}
\sum_{j}e^{i\mathbf{k}_{1}\mathbf{\cdot r}_{j}^{0}}\delta (\mathbf{r}-%
\mathbf{r}_{j}(t))=N\sum_{m}e^{i(\mathbf{k}_{1}+\mathbf{K}_{m})\cdot \left[
\mathbf{r-}\hat{n}f(t)\right] }.  \label{s0}
\end{equation}%
This sum enters the right hand side of Eq. (\ref{s1}). In the Fourier series
(\ref{s0}) we are interested in terms that are in resonance with the left
hand side of Eq. (\ref{s1}). For simplicity we assume that only two vectors,
namely, $\mathbf{k}_{1}$ and $\mathbf{k}_{2}=\mathbf{k}_{1}+\mathbf{K}_{b}$
have absolute values close to $\omega _{ab}/c$, where $\mathbf{K}_{b}$ is a
reciprocal lattice vector, see Fig. \ref{Reciprocal lattice}. The other
terms in (\ref{s0}) are off resonance and thus can be disregarded.
Therefore, one can write approximately%
\begin{align}
& \sum_{j}e^{i\mathbf{k}_{1}\mathbf{\cdot r}_{j}^{0}}\delta \left( \mathbf{r}%
-\mathbf{r}_{j}(t)\right)  \notag \\
\approx & Ne^{-i\mathbf{k}_{1}\cdot \hat{n}f(t)}e^{i\mathbf{k}_{1}\cdot
\mathbf{r}}+Ne^{-i\mathbf{k}_{2}\cdot \hat{n}f(t)}e^{i\mathbf{k}_{2}\cdot
\mathbf{r}}\,.  \label{f1}
\end{align}%
This approximation implies that the incident wave $\mathbf{k}_{1}$ is
coupled only with one Bragg wave $\mathbf{k}_{2}$.

Equation (\ref{f1}) suggests that one can look for a solution for $\Omega (t,\mathbf{r})$ in
the form of a superposition of these coupled waves%
\begin{equation}
\Omega (t,\mathbf{r})=\Omega _{1}(t)e^{i\mathbf{k}_{1}\cdot \mathbf{r}%
}+\Omega _{2}(t)e^{i\mathbf{k}_{2}\cdot \mathbf{r}}.
\end{equation}%
Then Eqs. (\ref{s1}) and (\ref{x0}) yield the following equations for $%
\Omega _{1}(t)$, $\Omega _{2}(t)$ and $\rho (t)$ (we take into account that $%
e^{i\mathbf{k}_{1}\mathbf{\cdot r}_{j}^{0}}=e^{i\mathbf{k}_{2}\mathbf{\cdot r%
}_{j}^{0}}$)
\begin{align}
& \left( \frac{\partial }{\partial t}+i\Delta _{1}\right) \Omega
_{1}(t)=i\Omega _{a}^{2}e^{-i\mathbf{k}_{1}\cdot \hat{n}f(t)}\rho (t)\,,
\label{n1} \\
& \left( \frac{\partial }{\partial t}+i\Delta _{2}\right) \Omega
_{2}(t)=i\Omega _{a}^{2}e^{-i\mathbf{k}_{2}\cdot \hat{n}f(t)}\rho (t)\,,
\label{n2} \\
& \frac{\partial \rho (t)}{\partial t}=i\Omega _{1}(t)e^{i\mathbf{k}%
_{1}\cdot \hat{n}f(t)}+i\Omega _{2}(t)e^{i\mathbf{k}_{2}\cdot \hat{n}f(t)}\,,
\label{n3}
\end{align}%
where
\begin{equation}
\Delta _{1,2}=\frac{c^{2}k_{1,2}^{2}-\omega _{ab}^{2}}{2\omega _{ab}}\approx
\nu _{1,2}-\omega _{ab}
\end{equation}%
are detunings of the two coupled waves from the nuclear transition frequency
$\omega _{ab}$. Taking the time derivative of both sides of Eqs. (\ref{n1}) and (%
\ref{n2}) and using Eq. (\ref{n3}), we obtain evolution equations for two $%
\gamma $-ray waves $\Omega _{1}(t)$ and $\Omega _{2}(t)$:%
\begin{align}
\bigg( & \frac{\partial }{\partial t}+i\mathbf{k}_{1}\cdot \hat{n}\dot{f}%
\bigg) \left( \frac{\partial }{\partial t}+i\Delta _{1}\right) \Omega _{1}
\notag \\
& +\Omega _{a}^{2}\left[ \Omega _{1}+\Omega _{2}e^{-i(\mathbf{k}_{1}-\mathbf{%
k}_{2})\cdot \hat{n}f(t)}\right] =0\,,  \label{x12} \\
\bigg( & \frac{\partial }{\partial t}+i\mathbf{k}_{2}\cdot \hat{n}\dot{f}%
\bigg) \left( \frac{\partial }{\partial t}+i\Delta _{2}\right) \Omega _{2}
\notag \\
& +\Omega _{a}^{2}\left[ \Omega _{2}+\Omega _{1}e^{i(\mathbf{k}_{1}-\mathbf{k%
}_{2})\cdot \hat{n}f(t)}\right] =0\,.  \label{x13}
\end{align}

Equations (\ref{x12}) and (\ref{x13}) constitute one of our main findings. These
equations describe two coupled harmonic oscillators whose parameters
periodically change in time. The varying of the parameters drives the
system. Namely, nuclei vibrations modulate coupling between two oscillators
as indicated by the $\Omega _{a}^{2}e^{\pm i(\mathbf{k}_{1}-\mathbf{k}%
_{2})\cdot \hat{n}f(t)}$ terms and, in addition, they periodically modulate the
oscillator's frequency by means of the Doppler shift $i\mathbf{k}_{1,2}\cdot
\hat{n}\dot{f}$.

It is known that parametric oscillators can have parametric resonances when
system's parameters are periodically modulated which can lead to
exponentially growing oscillations. An interesting question appears in this
context: Can Eqs. (\ref{x12}) and (\ref{x13}) yield exponentially growing
solutions which would imply that the high-frequency $\gamma $-ray field is
being generated at the expense of the energy stored in the low-frequency
nuclear vibrations? In the Appendix we show that the answer to this question
is that nuclear vibrations can not
excite nuclear transitions in the present model. Specifically, we show that
the sum of the energy of the high frequency field $\Omega (t,\mathbf{r})$
and that stored in the nuclear excitation is conserved no matter how nuclei
move.

Nevertheless, parametric resonance can be useful in the present problem.
Namely, it can substantially speed up energy transfer from one coupled
oscillator to another, that is, from the incident $\gamma $-ray beam to the
deflected one. This mechanism can be used to control propagation of $\gamma $ rays on a short time scale, which we discuss next.

\section{Beam deflection by coherent lattice vibration}

\label{sec:Beam refraction}

\subsection{Deflection by static lattice}

First we consider the interaction between the $\gamma $-ray field and a static nuclear
array. In this case there is no nuclear motion, so $f=0$ and Eqs. (\ref%
{n1})$-$(\ref{n3}) can be solved analytically. In particular, if $\Delta
_{1}=\Delta _{2}=\Delta $ the solution satisfying the initial condition $\Omega
_{1}(0)=A$, $\Omega _{2}(0)=0$ and $\rho (0)=0$ reads%
\begin{align}
\Omega _{1}(t)&=\frac{Ae^{-i\Delta t}}{2}\left( \frac{\omega _{+}e^{i\omega
_{-}t}-\omega _{-}e^{i\omega _{+}t}}{\sqrt{\Delta ^{2}+8\Omega _{a}^{2}}}+1%
\right) \,,  \label{x14} \\
\Omega _{2}(t)&=\frac{Ae^{-i\Delta t}}{2}\left( \frac{\omega _{+}e^{i\omega
_{-}t}-\omega _{-}e^{i\omega _{+}t}}{\sqrt{\Delta ^{2}+8\Omega _{a}^{2}}}-1%
\right) \,,  \label{x15} \\ 
\rho (t)&=-\frac{Ae^{-i\Delta t}}{\sqrt{\Delta ^{2}+8\Omega _{a}^{2}}}%
(e^{i\omega _{-}t}-e^{i\omega _{+}t})\,,  \label{x16}
\end{align}
where
\begin{equation}
\omega _{\pm }=\frac{1}{2}\left( \Delta \pm \sqrt{\Delta ^{2}+8\Omega
_{a}^{2}}\right) .  \label{x18}
\end{equation}

Equations (\ref{x14})$-$(\ref{x16}) yield that on resonance ($\Delta =0$)
\begin{align}
\Omega _{1}&=A\cos ^{2}(\Omega _{a}t/\sqrt{2})\,, \\
\Omega _{2}&=-A\sin ^{2}(\Omega _{a}t/\sqrt{2})\,, \\
\rho &=\frac{iA}{\sqrt{2}\Omega _{a}}\sin (\sqrt{2}\Omega _{a}t)\,.  \label{x17}
\end{align}%
That is, energy is periodically transferred back and forth between two
coupled waves on a time scale given by the collective nuclear frequency $%
\Omega _{a}$ which is proportional to the square root of the nuclear
density. Typically $1/\Omega _{a}$ is of the order of picoseconds.
Amplitudes of the $\gamma $-ray beams undergo collective oscillations \cite%
{Svid08a,Svid12} with frequency $\sqrt{2}\Omega _{a}$, as shown in Fig. \ref%
{Omega12OnOff}a.

According to Eq. (\ref{x17}), nuclei become excited during the energy
transfer between two $\gamma $ waves. Namely, the incoming $\gamma $ wave is
partially absorbed by the nuclear array. Absorption is followed by the
superradiant spontaneous emission of photons into the coupled wave.

\begin{figure}[h]
\begin{center}
\epsfig{figure=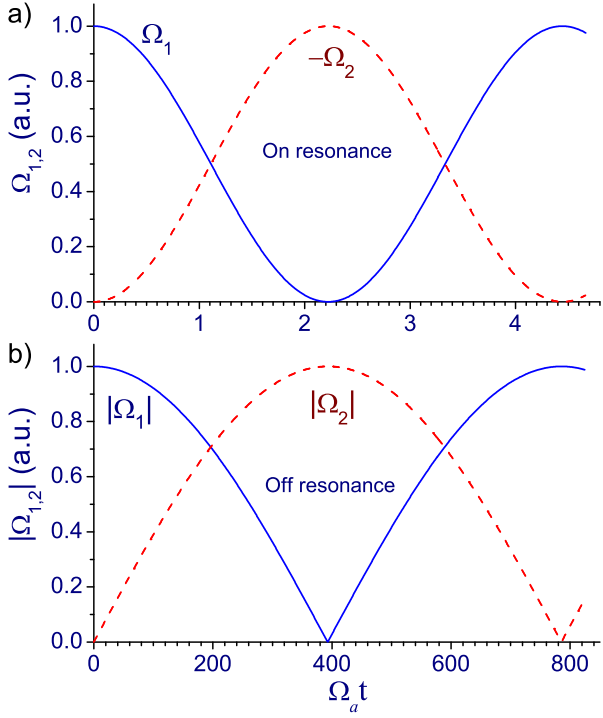, width=8cm}
\end{center}
\caption{(Color online) Time evolution of coupled $\protect\gamma $-ray
waves $\Omega _{1}(t)$ and $\Omega _{2}(t)$ produced by interaction with
static nuclear array. Initially $\Omega _{2}(0)=0$ and nuclei are in the
ground state. The solid line represents $|\Omega _{1}(t)|$ and the dashed line
indicates $|\Omega _{2}(t)|$. (a) The wave frequency is in resonance with the
nuclear transition. Energy is transferred back and forth between $\Omega
_{1}(t)$ and $\Omega _{2}(t)$ with collective frequency $\protect\sqrt{2}%
\Omega _{a}$ which typically lies in the terahertz range. (b) Off-resonance
interaction with frequency detuning $\Delta =250\Omega _{a}$. The energy
transfer occurs on a much longer time scale $\protect\pi\Delta /2\Omega
_{a}^{2}$.}
\label{Omega12OnOff}
\end{figure}

In contrast, if the wave frequency is off resonance, i.e. $\Delta \gg
\Omega _{a}$, the energy transfer between beams $\Omega _{1}$ and $\Omega
_{2}$ occurs over a much longer time%
\begin{equation}
t_{\text{tr}}^{0}=\frac{\pi }{|\omega _{-}|}\approx \frac{\pi |\Delta |}{%
2\Omega _{a}^{2}},
\end{equation}%
as shown in Fig. \ref{Omega12OnOff}(b).

One should mention that energy oscillations between two $\gamma $-ray modes,
referred to as the temporal Pendell\"{o}sung effect due to different hyperfine
transition frequencies at different nuclear sites, have been discussed in
\cite{Hannon89}. In Ref. \cite{Shvydko94} the Bragg switching of the $\gamma $%
-ray beam was realized using manipulation of nuclear spin states. In our
mechanism, oscillations appear due to the collective nature of the interaction
between light and the nuclear ensemble. Next we investigate how nuclear motion
affects energy transfer between the two coupled $\gamma $-ray waves.

\subsection{Beam deflection by oscillating lattice}

Here we assume that the nuclear array coherently vibrates with frequency $\nu
_{d}$ and the consider transformation of the incoming wave $\mathbf{k}_{1}$ into
the deflected wave $\mathbf{k}_{2}$. We assume that both waves are equally
detuned from the nuclear transition, that is $\Delta _{1}=\Delta _{2}=\Delta
$. We also assume that nuclei vibrate with amplitude $d$ along the direction
$\hat{n}$ perpendicular to $\mathbf{k}_{1}-\mathbf{k}_{2}$, as indicated in
Fig. \ref{Reciprocal lattice}. Then $\mathbf{k}_{1}\cdot \hat{n}=\mathbf{k}%
_{2}\cdot \hat{n}$. Introducing the dimensionless modulation amplitude%
\begin{equation}
\kappa =d\mathbf{k}_{1}\cdot \hat{n}
\end{equation}%
Eqs. (\ref{x12}) and (\ref{x13}) reduce to

\begin{align}
& \ddot{\Omega}_{1}+i(\Delta +F)\dot{\Omega}_{1}+(\Omega _{a}^{2}-\Delta
F)\Omega _{1}+\Omega _{a}^{2}\Omega _{2}=0\,,  \label{Omega1_1} \\
& \ddot{\Omega}_{2}+i(\Delta +F)\dot{\Omega}_{2}+(\Omega _{a}^{2}-\Delta
F)\Omega _{2}+\Omega _{a}^{2}\Omega _{1}=0,  \label{Omega2_1}
\end{align}%
where%
\begin{equation}
F(t)=\kappa \nu _{d}\cos (\nu _{d}t)
\end{equation}%
is a function that describes the modulation produced by nuclear motion.
The amplitude of the nuclei vibrations $d$ is much smaller than the spacing $a$
between nuclei. However, since the wave length of the nuclear transition is
also small compared to $a$ the modulation amplitude $\kappa $ could be of
the order of $1$.

The initial conditions for Eqs. (\ref{Omega1_1}) and (\ref{Omega2_1}) are $%
\Omega _{1}(0)=A$ and $\Omega _{2}(0)=0$. We assume that initially there is
no nuclear excitation [$\rho (0)=0$], which, according to Eqs. (\ref{n1})
and (\ref{n2}), yield $\dot{\Omega}_{1}(0)=-i\Delta A$ and $\dot{\Omega}%
_{2}(0)=0$.

Equations (\ref{Omega1_1}) and (\ref{Omega2_1}) have the integral of
motion
\begin{equation}
\Omega _{1}=\Omega _{2}+Ae^{-i\Delta t}.  \label{px1}
\end{equation}%
Plugging this into Eq. (\ref{Omega2_1}) and introducing $\tilde{\Omega}_{2}$
according to%
\begin{equation}
\Omega _{2}=e^{-i\Delta t}\left( \tilde{\Omega}_{2}-\frac{A}{2}\right)
\label{px2}
\end{equation}%
we obtain the following equation for $\tilde{\Omega}_{2}$
\begin{equation}
\frac{d^{2}\tilde{\Omega}_{2}}{dt^{2}}+i(F-\Delta )\frac{d\tilde{\Omega}_{2}%
}{dt}+2\Omega _{a}^{2}\tilde{\Omega}_{2}=0  \label{p2}
\end{equation}%
which is an equation of the parametric oscillator. Equation (\ref{p2}) has a solution
in terms of special functions, however, such a solution is not very insightful.
Instead, we derive an approximate solution that clearly shows the physics
behind the parametric speed up of the energy transfer. Introducing the function $%
u(t) $%
\begin{equation}
\frac{d\tilde{\Omega}_{2}}{dt}=\exp\bigg(-i\int_{0}^{t}(F(t^\prime)-\Delta )dt^{\prime }\bigg)u(t)\,,
\label{p3}
\end{equation}%
one can rewrite Eq. (\ref{p2}) as
\begin{equation}
\frac{du}{dt}=-2\Omega _{a}^{2}\exp\bigg(i\int_{0}^{t}(F(t^\prime)-\Delta )dt^{\prime }\bigg)
\tilde{\Omega}_{2}.  \label{p4}
\end{equation}%
Next we expand the exponential factor into the Fourier series%
\begin{align}
\exp\bigg(i\int_{0}^{t}(F(t^\prime)-\Delta )dt^{\prime }&\bigg)=e^{-i\Delta t}e^{i\kappa \sin (\nu
_{d}t)} \notag \\
=&e^{-i\Delta t}[ J_{0}(\kappa )+2iJ_{1}(\kappa )\sin (\nu
_{d}t) \notag \\
&+2J_{2}(\kappa )\cos (2\nu _{d}t)+\ldots ] ,  \label{p5}
\end{align}%
where $J_{n}(\kappa )$ are the Bessel functions. We assume that $\nu _{d}$
is close to $\Delta $ while $\tilde{\Omega}_{2}$ and $u$ are slowly varying
functions of time on the scale $1/\nu _{d}$. Then in the Fourier expansion (%
\ref{p5}) one can keep only the slowly varying term and approximately write%
\begin{equation}
\exp\bigg(i\int_{0}^{t}(F(t^\prime)-\Delta )dt^{\prime }\bigg)\approx J_{1}(\kappa )e^{i\left( \nu
_{d}-\Delta \right) t}.
\end{equation}%
As a result, Eqs. (\ref{p3}) and (\ref{p4}) reduce to
\begin{align}
&\frac{d\tilde{\Omega}_{2}}{dt}=J_{1}(\kappa )e^{-i\left( \nu _{d}-\Delta
\right) t}u\,, \\
&\frac{du}{dt}=-2J_{1}(\kappa )\Omega _{a}^{2}e^{i\left( \nu _{d}-\Delta
\right) t}\tilde{\Omega}_{2}\,,
\end{align}
which can be solved analytically. Plugging this solution into Eqs. (\ref{px1})
and (\ref{px2}) we finally obtain%
\begin{align}
\Omega _{1}&=\frac{Ae^{-i\Delta t}}{2}\left( \frac{\omega _{+}e^{-i\omega
_{-}t}-\omega _{-}e^{-i\omega _{+}t}}{\sqrt{\left( \nu _{d}-\Delta \right)
^{2}+8J_{1}^{2}(\kappa )\Omega _{a}^{2}}}+1\right)\,, \\
\Omega _{2}&=\frac{Ae^{-i\Delta t}}{2}\left( \frac{\omega _{+}e^{-i\omega
_{-}t}-\omega _{-}e^{-i\omega _{+}t}}{\sqrt{\left( \nu _{d}-\Delta \right)
^{2}+8J_{1}^{2}(\kappa )\Omega _{a}^{2}}}-1\right)\,, \\
\end{align}%
where
\begin{equation}
\omega _{\pm }=\frac{1}{2}\left( \nu _{d}-\Delta \pm \sqrt{\left( \nu
_{d}-\Delta \right) ^{2}+8J_{1}^{2}(\kappa )\Omega _{a}^{2}}\right) .
\end{equation}%
When $\nu _{d}=\Delta $ we find%
\begin{align}
\Omega _{1}&=Ae^{-i\Delta t}\cos ^{2}\left( \frac{J_{1}(\kappa )}{\sqrt{2}}%
\Omega _{a}t\right)\,,  \label{pp1} \\
\Omega _{2}&=-Ae^{-i\Delta t}\sin ^{2}\left( \frac{J_{1}(\kappa )}{\sqrt{2}}%
\Omega _{a}t\right)\,.  \label{pp2}
\end{align}%
Eqs. (\ref{pp1}) and (\ref{pp2}) show that the rate of energy transfer
between two coupled waves depends on the amplitude of the nuclear
vibrations. The optimum value of the modulation amplitude $\kappa $
corresponds to maximum of $J_{1}(\kappa )$, that is $\kappa =1.841$ which
gives $J_{1}(\kappa )/\sqrt{2}=0.411$. For larger $\kappa $ the transfer
rate oscillates following $J_{1}(\kappa )$.

For $\kappa \ll 1$ one can use the expansion $J_{1}(\kappa )\approx \kappa /2$.
Then Eq. (\ref{pp2}) yields that energy transfer time between two waves is

\begin{equation}
t_{\text{tr}}=\frac{\sqrt{2}\pi }{\kappa \Omega _{a}}.  \label{pp3}
\end{equation}

\begin{figure}[h]
\begin{center}
\epsfig{figure=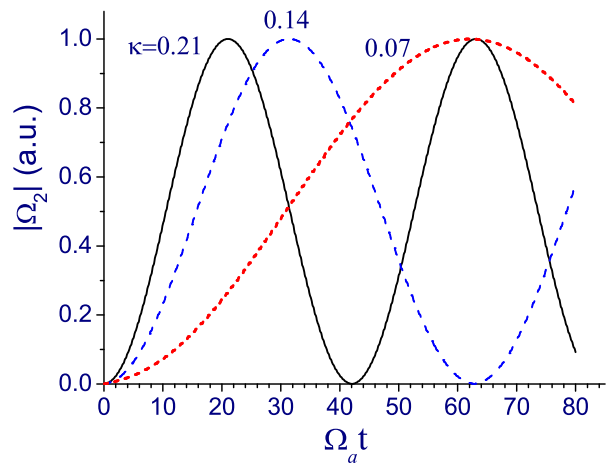, width=8cm}
\end{center}
\caption{(Color online) Time evolution of the deflected $\protect\gamma $%
-wave $|\Omega _{2}(t)|$ obtained by numerical solution of Eqs. (\protect\ref%
{Omega1_1}) and (\protect\ref{Omega2_1}) for different values of the
modulation amplitude $\protect\kappa =0.21$, $0.14$ and $0.07 $. In
simulations we set $\protect\nu _{d}=\Delta =250\Omega _{a}$.
The transformation time between two $\protect\gamma $-waves $\Omega _{1}$ and $%
\Omega _{2}$ is $t_{\text{tr}}\approx \protect\sqrt{2}\protect\pi /\protect%
\kappa \Omega _{a}$. }
\label{Omega2_k}
\end{figure}

In Fig. \ref{Omega2_k} we plot $\Omega _{2}(t)$ for different values of the
modulation amplitude $\kappa $ obtained by numerical solution of Eqs. (\ref%
{Omega1_1}) and (\ref{Omega2_1}). Our analytical result (\ref{pp2}) gives
essentially the same curves.

When the incident $\gamma $ wave $\Omega _{1}$ is off resonance with the
nuclear transition, the time it takes for the energy to transfer from $%
\Omega _{1}$ into the deflected wave $\Omega _{2}$ can substantially vary
with or without nuclear vibrations. This can be used for fast switching of
the wave propagation that can be achieved in the regime $\kappa \Delta \gg
\Omega _{a}$. If the $\gamma $-wave detuning $\Delta $ is large enough then
the wave will pass through the static crystal without deflection. However,
if the nuclear vibrations are suddenly turned on with $\nu _{d}=\Delta $ the
incident $\gamma $ wave will be deflected on a time scale given by Eq. (\ref%
{pp3}), which could be a few picoseconds.

\begin{figure}[h]
\begin{center}
\epsfig{figure=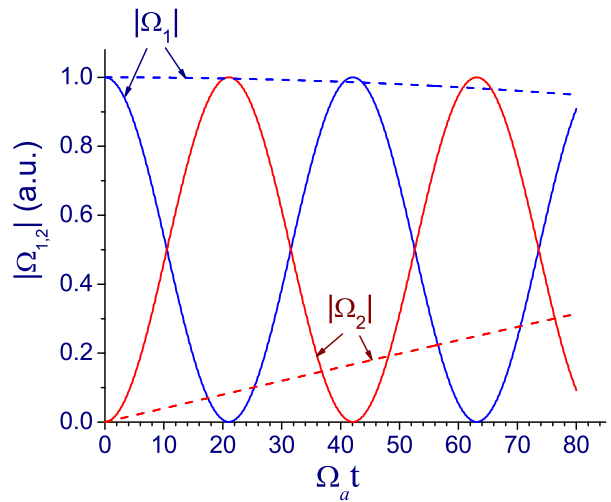, width=8cm}
\end{center}
\caption{(Color online) Illustration of the $\protect\gamma $-ray switch
operation. The incident $\protect\gamma $ wave $\Omega _{1}$ is detuned from
the nuclear transition by $\Delta =250\Omega _{a}$. The dashed lines show
the transformation of $\Omega _{1}$ into $\Omega _{2}$ for a static crystal and the solid lines are for nuclear array vibrating with frequency $\protect%
\nu _{d}=\Delta $ and modulation amplitude $\protect\kappa =0.21$. }
\label{Omega12_compare}
\end{figure}

Figure \ref{Omega12_compare} demonstrates the effect for a medium with $\Omega
_{a}=0.8$ THz assuming that the incident wave is detuned from the nuclear
transition by $\Delta =250\Omega _{a}$. For a static crystal the fields $%
\Omega _{1}$ and $\Omega _{2}$ are shown by dashed lines. Without nuclear
vibrations it takes $t_{\text{tr}}^{0}=491$~ps for the wave $\Omega _{1}$ to
convert into $\Omega _{2}$. If the crystal size is smaller than $ct_{\text{tr%
}}^{0}=15$ cm the incident wave passes through. However, if the nuclear
array vibrates with modulation amplitude $\kappa =0.21$ the conversion time
becomes $t_{\text{tr}}=26$~ps and, thus, the wave will be deflected at a
length of $0.8$ cm (solid lines in Fig. \ref{Omega12_compare}).

One should note that the $\gamma $-ray switch can also operate in the on
resonance regime. Namely, when the incident wave is on resonance with the
nuclear transition it converts fast into the deflected wave on a time scale $%
\pi /\sqrt{2}\Omega _{a}$. Turning on nuclear vibrations would destroy the
resonance interaction and make the wave pass through the crystal.

\section{Discussion}

The physics behind the speed up of the energy transfer between two waves can be
understood as a parametric resonance in a system of coupled oscillators. A
single oscillator whose frequency is periodically modulated provides a
simple example of parametric resonance. The motion of such an oscillator is
described by the Mathieu's equation%
\begin{equation}
\ddot{x}+\omega _{0}^{2}\left[ 1+\delta \cos (\nu _{d}t)\right] x=0,
\label{h1}
\end{equation}%
where $\delta $ is the modulation amplitude. If $\delta =0$ then the system has
two natural frequencies $\pm \omega _{0}$. If the system's parameters vary
with frequency $\nu _{d}$ equal to the difference between natural
frequencies, that is $\nu _{d}=2\omega _{0}$, the oscillator phase locks to
the parametric variation and undergoes a parametric resonance absorbing energy
at a rate proportional to the energy it already has.

A similar situation takes place if the system has several natural frequencies
(normal modes). To achieve parametric resonance the modulation frequency $%
\nu _{d}$ must match the difference between two normal mode frequencies~\cite%
{Hsu63}. This is known as the difference combination resonance \cite%
{Nayf2000}. In the present problem the natural frequencies of the coupled
light-nuclear system are determined from the solution (\ref{x14}) and (\ref%
{x15}) obtained for the static lattice. Thus, if the frequency of the
nuclear vibrations matches the frequency difference, namely, $\nu
_{d}=\omega _{+}-\omega _{-}=\sqrt{\Delta ^{2}+8\Omega _{a}^{2}}$ the system
undergoes parametric resonance, which speeds up the energy transfer between
two $\gamma $ waves. This phenomenon is analogous to parametric frequency
mixing in propagating circuits~\cite{Tien58}, in which power can flow back
and forth between the two coupled circuits if the coupling reactance
variation frequency matches their frequency difference.

Combination parametric resonance at the frequency difference between two
normal modes of the coupled light-atom system is the essence of the QASER~%
\cite{QASER}, a device that can generate high-frequency (e.g., XUV) coherent
light by driving an atomic medium with a low frequency (e.g., infrared) field
\cite{Svid13}. Contrary to the laser, the QASER does not require any atomic
population in the excited state and yields high-frequency light
amplification. In the case of the QASER the external field drives the atomic
transition which produces modulation of the atom-field coupling strength and
yields gain at high frequency. In the present model, unlike the QASER,
modulation is produced by the nuclear motion, which does not yield
amplification of the high-frequency ($\gamma$) field. However, parametric
resonance and collective effects of the light interaction with a nuclear array
enhance the rates of the radiation absorption and reemission. As a result,
energy transfer between two $\gamma $ waves occurs on a much shorter
superradiant time scale determined by the collective nuclear frequency $%
\Omega _{a}$ which is of the order of terahertz. The combination of the Dicke
superradiance, Bragg diffraction, and combination parametric resonance introduces
interesting features to our problem and allows us to achieve fast
manipulation of the $\gamma $-ray propagation.

If the incident $\gamma $ wave is far detuned from the nuclear transition by
the amount $\Delta \gg \Omega _{a}$ then the light-nuclear interaction is weak
and the $\gamma $ wave passes through the static nuclear array. We found
that if we make the nuclei vibrate coherently with frequency $\Delta $
then the combination parametric resonance effectively enhances the light-nuclear
interaction strength. As a result, the incident $\gamma $ wave undergoes
deflection into a wave propagating at the Bragg angle on a short
superradiant time scale. The maximum energy transfer rate is achieved for
the amplitude of nuclear oscillations $d\sim \lambda _{ab}/2\pi $, where $%
\lambda _{ab}$ is the wavelength of the nuclear transition. Since $\lambda
_{ab}$ is typically much smaller than the spacing between nuclei in crystals the
required nuclear vibrations are also small. Such nuclear motion can be
realized, e.g., in ionic crystals illuminated by a strong driving optical
laser field.

Our findings could be used for manipulation of the propagation direction of $%
\gamma $ rays on a picosecond time scale.

\begin{acknowledgments}
We thank O. Kocharovskaya for useful discussions. We gratefully acknowledge
support from the National Science Foundation through Grants No. PHY-1205868 and No. PHY-1241032
and the Robert A. Welch Foundation (Award No. A-1261).
\end{acknowledgments}

\appendix

\section{Conservation of high frequency energy component during light
interaction with vibrating nuclei}

We start from Eqs. (\ref{s1}) and (\ref{x0}) describing light propagation
through a moving crystal lattice%
\begin{align}
& \left( \frac{\partial }{\partial t}+\frac{c^{2}}{2i\omega _{ab}}\left[
\left( \frac{\omega _{ab}}{c}\right) ^{2}+\nabla ^{2}\right] \right) \Omega
(t,\mathbf{r})  \notag \\
& \qquad \qquad \qquad =i\frac{\Omega _{a}^{2}}{N}\sum_{j}\rho ^{j}(t)\delta
(\mathbf{r}-\mathbf{r}_{j}(t)),  \label{s1aa} \\
& \frac{\partial \rho ^{j}(t)}{\partial t}=i\Delta \omega _{ab}(t)\rho
^{j}(t)+i\Omega (t,\mathbf{r}_{j}(t))\,.  \label{x0aa}
\end{align}%
In Eq. (\ref{x0aa}) we introduced an additional time-dependent frequency
shift $\Delta \omega _{ab}(t)$ to include possible external modulation of
the nuclear transition frequency. Multiplying both sides of Eq. (\ref{s1aa})
by $\Omega ^{\ast }(t,\mathbf{r})$ we obtain%
\begin{align}
\Omega ^{\ast }(t,\mathbf{r})& \left( \frac{\partial }{\partial t}+\frac{%
c^{2}}{2i\omega _{ab}}\left[ \left( \frac{\omega _{ab}}{c}\right)
^{2}+\nabla ^{2}\right] \right) \Omega (t,\mathbf{r})  \notag \\
=& i\frac{\Omega _{a}^{2}}{N}\sum_{j}\rho ^{j}(t)\Omega ^{\ast }(t,\mathbf{r}%
_{j}(t))\delta \left( \mathbf{r}-\mathbf{r}_{j}(t)\right) \,.  \label{a1}
\end{align}%
While Eq. (\ref{x0aa}) yields%
\begin{equation}
i\rho ^{j}(t)\Omega ^{\ast }(t,\mathbf{r}_{j}(t))=-\rho ^{j}(t)\dot{\rho}%
^{j\ast }(t)-i\Delta \omega _{ab}(t)|\rho ^{j}(t)|^{2}\,.  \label{mid1}
\end{equation}%
Plugging Eq. (\ref{mid1}) into Eq. (\ref{a1}) gives%
\begin{align}
& \Omega ^{\ast }(t,\mathbf{r})\left( \frac{\partial }{\partial t}+\frac{%
c^{2}}{2i\omega _{ab}}\left[ \left( \frac{\omega _{ab}}{c}\right)
^{2}+\nabla ^{2}\right] \right) \Omega (t,\mathbf{r})  \notag \\
+& \frac{\Omega _{a}^{2}}{N}\sum_{j}\rho ^{j}(t)\dot{\rho}^{j\ast }(t)\delta
(\mathbf{r}-\mathbf{r}_{j}(t))  \notag \\
+& i\Delta \omega _{ab}(t)\frac{\Omega _{a}^{2}}{N}\sum_{j}|\rho
^{j}(t)|^{2}\delta (\mathbf{r}-\mathbf{r}_{j}(t))=0\,.  \label{mid2}
\end{align}%
Adding to Eq. (\ref{mid2}) its complex conjugate we obtain%
\begin{align}
\frac{\partial }{\partial t}|\Omega (t,\mathbf{r})|^{2}& +\frac{c^{2}}{%
2i\omega _{ab}}\nabla \left[ \Omega ^{\ast }(t,\mathbf{r})\nabla \Omega (t,%
\mathbf{r})-c.c\right]  \notag \\
& +\frac{\Omega _{a}^{2}}{N}\sum_{j}\delta \left( \mathbf{r}-\mathbf{r}%
_{j}(t)\right) \frac{\partial }{\partial t}|\rho ^{j}(t)|^{2}=0\,.
\label{mid3}
\end{align}%
Integrating Eq. (\ref{mid3}) over space and taking into account that for
weakly excited nuclei $|\rho ^{j}|^{2}=\rho _{aa}^{j}$, where $\rho
_{aa}^{j} $ is the excited state population of the nucleus $j$, we find%
\begin{equation}
\int d\mathbf{r}|\Omega (t,\mathbf{r})|^{2}+\frac{\Omega _{a}^{2}}{N}%
\sum_{j}\rho _{aa}^{j}(t)=\text{const}\,.  \label{ec}
\end{equation}%
Equation (\ref{ec}) shows that the sum of the energy of the high frequency field $%
\Omega (t,\mathbf{r})$ and that stored in nuclear excitation is conserved
no matter how nuclei move. This implies that nuclear motion can not amplify
high-frequency field.

\end{document}